\documentclass[twocolumn,showpacs,preprintnumbers,amsmath,amssymb]{revtex4}

\usepackage{graphicx}
\usepackage{dcolumn}
\usepackage{bm}
\usepackage{epsfig}
\usepackage{natbib}

\begin{document}
\preprint{APS}

\title{Ground state heavy baryon production in a relativistic quark-diquark model}

\author{M.A. Gomshi Nobary}\email{mnobary@razi.ac.ir}
\affiliation{Department of Physics, Faculty of Science, Razi
University, Kermanshah, Iran.}
\author{R. Sepahvand}\email{re_sepahvand@yahoo.com}%
\affiliation{Department of Physics, Faculty of Science, Lorestan
University, Khoramabad, Iran.}

\date{\today}

\begin{abstract}
We use current-current interaction to calculate the fragmentation
functions to describe the production of spin-1/2, spin-1/2$'$ and
spin-3/2 baryons with massive constituents in a relativistic
quark-diquark model. Our results are in their analytic forms and
are applicable for singly, doubly and triply heavy baryons. We
discuss the production of $\Omega_{bbc}$, $\Omega_{bcc}$ and
$\Omega_{ccc}$ baryons in some detail. The results are
satisfactorily compared with those obtained for triply heavy
baryons calculated in a perturbative regime within reasonable
values of the parameters involved.
\end{abstract}

\maketitle
\section {Introduction}
Heavy baryon physics is one of the important topics in recent
particle physics. One of the reasons is the improvement of the
experimental data [1]. Another reason rests on the theoretical
side where valuable phenomenological aspects of treating such
states have been improved. Quantum chromodynamics serves as a
leading candidate to study the production and decay properties of
such states. Heavy quark effective theory [2] has also proved to
be a useful tool in situations where heavy quark and light
degree(s) of freedom both are present, namely, for singly and
doubly heavy baryons. It has proved to be successful in some
circumstances specially where the heavy quark mass could be
approximated infinity with respect to mass scale in the process
[3]. Heavy baryons have also been studied in relativistic quark
models mainly in favor of singly and doubly heavy baryons [4].
Another valuable idea is the application of diquarks [5].
Undoubtedly it reduces the task of treating a three body system to
a two body system. Its application in the case of light quarks
have proved to be successful in most of the situations.

Description of hadronic properties is accomplished by the set of
Dyson-Schwinger equations [6]. The simplest hadrons are mesons
which are color singlet bound states of a quark and an anti-quark.
They are described by solutions of the homogenous Bethe-Salpeter
equation (BSE) for $\overline q q$ states. The Bethe-Salpeter
amplitudes of different types of mesons such as pseudo-scalar,
vector, etc. are characterized by different Dirac structures. In
addition to $\overline q q$ bound states, there are $qq$ states by
studying the corresponding BSE. Single gluon exchange leads to an
interaction that is attractive for diquarks in a color antitriplet
configuration. Two quarks can be coupled in either a color sextet
or a color antitriplet state. Furthermore, it is the diquark in a
color antitriplet state that can couple with a quark to form a
color-singlet baryon. Therefore only the antitriplet configuration
of $qq$ states are considered here. Similar to the case of mesons,
the different types of diquarks are characterized by different
Dirac structures.

On the other hand we have an interesting situation regarding heavy
baryons in which different combinations of light and heavy flavor
give raise to singly, doubly and triply heavy states. Wide variety
of speculations have been applied to understand the production and
decay properties of these states. It seems that it would be
interesting to consider a case in which all possibilities are
involved benefiting the idea of diquark. In combining a quark with
a diquark, the state of the diquark will play an important role.
While the combination of a scalar diquark with a heavy quark will
end up with a spin-1/2 baryon, the similar situation with a vector
diquark will produce either a spin-1/2 (generally called as
spin-1/2$'$) or a spin-3/2 baryon. Such a scenario is to be
studied in more detail here where the constituents are assumed
massive and relativistic. We will focus our attention to the
production of such states, and obtain three fragmentation
functions to describe the fragmentation production of them and
compare our results with available theoretical results.

Our strategy is as follows. We discuss the heavy flavor diquarks
in section II. In section III we specify the spin wave functions
for baryons in their ground and in possible spin states. The
section IV is devoted to obtain the fragmentation functions and we
illustrate the application of our results in the case of
$\Omega_{bbc}$, $\Omega_{bcc}$ and $\Omega_{ccc}$ baryons in
section V where we also discuss our results.

\section{Heavy flavor diquarks }
Application of diquark model is achieved by calculating the
Feynman diagrams with rules for point-like particles. To embed the
composite nature of a diquark, phenomenological vertex functions
should be introduced. Parameterization of the 3-point functions
and diquark form factors is obtained from the requirement that,
asymptotically the diquark model evolves into the pure quark
model. Moreover that there is no direct information about
chromomagnetic form factors. We may expect that the ordinary
electromagnetic form factors will have the same functional form as
their QCD counterparts since the source of both form factors is
the matrix elements of a conserved vector operator. Here the
vector operator is the color octet gluon field. The
parameterization of the diquark form factors may be inferred from
quark-diquark models of the nucleons. In this view the relevant
form factors in the space-like region are written as [7]
\begin{eqnarray}
F_{S}(Q^2)&=&\biggl(1+\frac{Q^2}{Q^2_{S}}\biggr)^{-1},\quad
F_{E}(Q^2)=\biggl(1+\frac{Q^2}{Q^2_{V}}\biggr)^{-2},\nonumber\\
F_M(Q^2)&=&(1+\kappa )F_{E}(Q^2),\qquad F_Q(Q^2)=0.
\end{eqnarray}
It is assumed that the scalar and vector diquark form factors to
have simple and dipole forms respectively with pole positions at
$Q_S$ and $Q_V$ above 1 GeV [8]. In the above, $\kappa$ is the
anomalous chromomagnetic dipole moment. The form factor related to
chromoelectric quadrupole moment is set equal to zero [9]. In the
case of heavy diquarks, the following expressions for the above
form factors are also proposed

\begin{eqnarray}
F_{S}(Q^2)&=&\frac{Q^2_{S}}{Q^2},\qquad
F_{E}(Q^2)=\biggl(\frac{Q^2_{V}}{Q^2}\biggr)^{2},\nonumber\\
F_M(Q^2)&=&(1+\kappa )F_{E}(Q^2),\qquad F_Q(Q^2)=0.
\end{eqnarray}
The above form factors are more consistent where heavy flavor is
involved in heavy baryon production. The parameters $Q_S$ and
$Q_V$ are free parameters and have a crucial role in determination
of the fragmentation probabilities. We will comment more about the
values of $Q_S$ and $Q_V$ in the last section.

 The gluon-diqurk coupling will be different for scalar and vector
diquarks. The form of these couplings are determined by the
four-momentum flow and polarization four vectors in the Feynman
diagram [10]. The Feynman rule for the coupling of a gluon to a
scalar diquark (SgS-vertex) reads as

\begin{eqnarray}
ig_s T^a_{ij}(k+k'),
\end{eqnarray}
and for similar coupling for a vector diquark (VgV-vertex), we
have

\begin{eqnarray}
ig_s
T^a_{ij}\bigl\{g_{\alpha\beta}(k+k')_\mu&-&g_{\mu\alpha}\big[(1+\kappa)k-\kappa
k')\big]_\beta \nonumber\\&-&g_{\mu\beta}\bigl[(1+\kappa)k'-\kappa
k\bigr]_\alpha\big\}.
\end{eqnarray}
Here $k$ and $k'$ are four momenta for diquark and anti-diquark
respectively.

\section{Ground state spin wave functions for heavy diquarks and heavy baryons}

In this section we try to investigate the the way in which a heavy
quark and a heavy diquark combine together to for a heavy baryon.
it is assumed that the diquark and the produced baryon are in
their ground states. We emphasis that while the ground state
diquark composed of different flavors is a scalar spin-0 state, a
diquark with identical constituents should be considered in a
vector spin-1 state due to Fermi statistics.

In the quark-diquark model, ground state heavy baryons are
composed of a heavy quark $Q$ and a heavy diquark system $D$ with
spin-0 or spin-1, moving in a S-wave state. We denote the spin
wave functions of a heavy spin-0 diquark by $\chi$ and a spin-1
diquark by $\chi^\mu$. These functions are normalized according to
[3]

\begin{eqnarray}
(\chi,\chi)=1, \qquad (\chi^\mu,\chi^\nu)=-g^{\mu\nu}+\frac{k^\mu
k^\nu}{m_D^2}.
\end{eqnarray}
It is sometimes convenient to transform to the spherical basis for
the spin-1 diquark which can be done with the help of the spin-1
polarization vector. For $\lambda=\pm 1,\;0$, we may write
\begin{eqnarray}
\chi(1,\lambda)=\epsilon_\mu(\lambda)\chi^\mu,
\end{eqnarray}
and the inverse
\begin{eqnarray}
\chi^\mu=\sum_\lambda\epsilon^{*\mu}(\lambda)\chi(1,\;\lambda),
\end{eqnarray}
where the polarization four-vector for different helicity states
are defined as usual
\begin{eqnarray}
\epsilon_{\mu}(\pm)&=&\mp\frac{1}{\sqrt{2}}(0,\;1,\;\pm
i,0),\nonumber\\
\epsilon_{\mu}(0)&=&(|{\bf k}|,\;0,\;0,\;k_\circ).
\end{eqnarray}

Since the spin wave functions $\chi$ and $\chi^\mu$ satisfy the
Bargmann-Wigner equation on both spinor labels, they are spin wave
functions built from constituent on-mass shell quarks [3].

In combining a heavy quark and a diquark, one obtains a ground
state heavy baryon. The ground state diquark is either in spin-0
or in spin-1 state. Therefore there will be different
possibilities for the spin state of the heavy baryon. These
possibilities are shown in Table I. The spin wave function
$\Psi_{\alpha\beta\gamma}$ of a ground state heavy baryon is
written down by invariant coupling between the diquark spin
tensors $\chi$ and $\chi^\mu$ and the heavy quark spinor tensors
$u$ and $u^\mu$. The heavy quark spinor tensors $u$ and $u^\mu$
involve the heavy baryon spinor $U_B$ and the Rarita-Schwinger
spinor vector $\Psi^\mu$. We then may write [3]

\begin{widetext}
\begin{eqnarray}
{\rm
Scalar\;diquark;\;spin-1/2\;baryon}:\;\Psi_{\alpha\beta\gamma}&=&\chi_{\alpha\beta}u_\gamma\equiv\chi
U_B,\nonumber\\
{\rm
Vector\;diquark;\;spin-1/2'\;baryon}:\;\Psi_{\alpha\beta\gamma}&=&\chi_{\alpha\beta}^\mu
u_{\mu,\gamma}\equiv\chi^\mu\Big\{\frac{1}{\sqrt{3}}\gamma^\mu\gamma_5
U_B\Big\},\nonumber\\
{\rm
Vector\;diquark;\;spin-3/2\;baryon}:\;\Psi_{\alpha\beta\gamma}&=&\chi_{\alpha\beta}^\mu
u_{\mu,\gamma}\equiv\chi^\mu\Psi_\mu.
\end{eqnarray}
\end{widetext}

In the constituent quark model, the explicit forms of a scalar and
a vector diquark spin wave functions are respectively given by
[11]

\begin{eqnarray}
\chi&=&\frac{1}{2\sqrt{2}m_D}\bigl[(\rlap/
k+m_D)\gamma_5\bigr],\nonumber\\
\chi^\mu&=&\frac{1}{2\sqrt{2}m_D}\bigl[(\rlap/
k+m_D)\gamma^\mu\bigr].
\end{eqnarray}
Therefore the spin wave functions for the heavy baryons may be
constructed from these ingredients.

\begin{table}
\caption{The ground state triply heavy baryons with corresponding
possible spin states when considered to be formed in quark-diquark
model. The relevant diquark spin states are also shown. }\vskip
.5cm
\begin{ruledtabular}
\begin{tabular}{c c c}
 Baryon Formation Process&Diquark spin& Baryon spin\\ \hline
$c\rightarrow \Omega_{ccc},\;\Omega_{ccc}^*$ &$1$&$\frac{1}{2},\;\frac{3}{2}$  \\
$b\rightarrow \Omega_{bbb},\;\Omega_{bbb}^*$&1&$\frac{1}{2},\;\frac{3}{2} $ \\
$b\rightarrow \Omega_{bbc}\;\qquad\;$&0&$\frac{1}{2} $\\
$c\rightarrow \Omega_{bbc},\;\Omega_{bbc}^*$&1&$\frac{1}{2},\;\frac{3}{2} $\\
$c\rightarrow \Omega_{bcc}\;\qquad\;$ &0&$\frac{1}{2} $ \\
$b\rightarrow \Omega_{bcc},\;\Omega_{bcc}^*$ &1&$\frac{1}{2},\;\frac{3}{2}$ \\
\end{tabular}
\end{ruledtabular}
\end{table}

\section{Fragmentation functions}

The formation of a baryon in a quark-diquark model is described in
Fig. 1 where the diquark which attaches to the initial state heavy
quark may be in a scalar or a vector state. Fragmentation is
usually described by the function $D(z,\mu_\circ)$, where $z$
being the energy-momentum fraction taken by the baryon and
$\mu_\circ$ is the scale at which such function is calculable in
perturbative QCD. This function may be put in the following form

\begin{figure}
\centering\epsfig{file=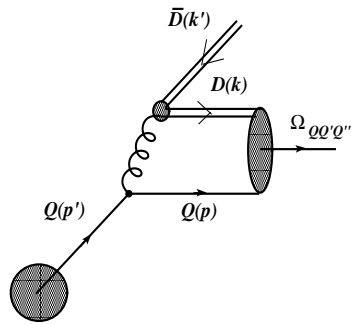,height=7cm,width=9.cm,clip=} \hspace{1.5cm} %
\vspace{-2.0cm} \caption{ The lowest order Feynman diagram
contributing to the fragmentation of a heavy quark ($Q$) into a
heavy baryon with three massive constituents ($\Omega_{QQ'Q''}$)
in a quark-diquark model. The four momenta are labelled. The
diquark $D$ consists of two heavy flavor $Q'$ and $Q''$ and
$\overline D$ of $\overline{Q'}$ and $\overline{Q''}$. }
\end{figure}

\begin{eqnarray}
D_{Q\rightarrow QQ'Q''}(z,\mu_\circ)&=& \frac{1}{2}\sum_{s}\int
d^3{\bf p}d^3{\bf k}d^3{\bf k'}|{ T_B}|^2\nonumber\\
&&\times\delta^3({\bf p}+{\bf k}+{\bf k'}-{\bf p'}),
\end{eqnarray}
where ${T_B}$ is the scattering amplitude for the fragmentation
process. The average over initial spin state and summation over
all final state particle spins are included for production of
unpolarized baryon state.

To obtain the hard scattering amplitude, we consider the currents
produced by the diquark pair and the initial state heavy quark
with appropriate coupling to a vector gluon. Such a current for a
scalar diquark may be put in the following form

\begin{eqnarray}
j^{\mu}_{S}\sim g_s F_{S}(Q^2)q^\mu \overline{\chi}'^(k').\chi(k)
e^{-iq.x},
\end{eqnarray}
where $k$ and $k'$ are the scalar diquark and anti-diquark
four-momenta with $q=k+k'$ and $F_S(Q^2)$ is the relevant form
factor given in (1). Here $Q^2=-q^2$. The tensors $\chi(k)$ and
$\chi'(k')$ are the spin wave functions for diquark and
anti-diquark respectively. Similarly the current for a vector
diquark with a coupling to a vector gluon casts into the following
[12]

\begin{widetext}
\begin{eqnarray}
j^{\mu}_{V}\sim g_s\Big\{ -F_E(Q^2)\Big[\chi (k).\overline{\chi}'
(k')\Big]q^\mu  +F_M(Q^2)\Big[(k.\chi' (k'))\overline{\chi}^{\mu
}+(k'.\overline{\chi}(k))\chi'^{\mu }\Big]\Big\}e^{-iq.x}.
\end{eqnarray}
\end{widetext} Here $F_E(Q^2)$ and $F_M(Q^2)$ are chromoelectric
and chromomagnetic form factors given in (1). The current produced
by the initial state heavy quark with a coupling to a vector gluon
may be written in the following form
\begin{eqnarray}
j_{\mu}^Q \sim g_s
\big[\bar{u}(p)\gamma_{\mu}u(p')\big]e^{-i(p-p').x}.
\end{eqnarray}

Now we write down the general form of the amplitudes for
production of a ground state heavy baryon with possible spin
states. In doing so first we form the hard scattering amplitude
using the currents (12), (13) an (14). Then we convolute the
result with the momentum space distribution amplitude to obtain
the total amplitude. The distribution amplitude is chosen to have
the following form

\begin{eqnarray}
\Phi_{B}=\frac{f_{B}}{m_B}\delta\Big(x_{i}-\frac{m_{i}}{m_{B}}\Big),
\end{eqnarray}
where $m_{B}$ and $f_{B}$ are mass and decay constant for the
baryon. With the distribution amplitude given by (15), it is
assumed that constituents of the baryon move almost collinearly
and that the effects of Fermi motion is disregarded. Eventually in
the first order perturbation theory the total amplitude for the
process in discussion, in a general case, will have the following
form

\begin{eqnarray}
T_{S,V}=-i\int
[dx]d^{4}x\Phi_B\Bigl\{j^{\mu}_{S,V}\Big(\frac{1}{q^2}\Big)j_{\mu}^Q\Bigr\}
,
\end{eqnarray}
where $[dx]=dx_{1}dx_{2}\delta(1-x_{1}-x_{2})$, $x_1$ and $x_2$
being the energy momentum ratios carried by the heavy quark and
the diquark. The currents $j^{\mu}_{S,V}$ and $j_{\mu}^Q$ are
given by (12), (13) and (14). We are now in a position to write
down the amplitudes corresponding to the possible attachments of
the initial state heavy quark with the diquark to form a ground
state baryon with a definite spin state. For a scalar diquark to
combine with a heavy quark giving rise to a spin-1/2 triply heavy
baryon, we have

\begin{widetext}
\begin{eqnarray}
T_{S{1/2}}=\frac{g_s^2C_{F}f_B }{8m_B^2\sqrt{2
p_{0}p'_{0}k_{0}k'_{0}}}\frac{F_S(Q^2)}{D_\circ q^2}\bigl\{
{\overline U}_B[\chi' (\rlap /P +m_B)\gamma_5(\rlap /k' +m_D)]
u(p')\bigr\}.
\end{eqnarray}
\end{widetext} In the above, the conservation of three momentum is
implicit. Since the initial state heavy quark is not on its mass
shell, we have energy non-conservation, and have performed the
energy integration to reproduce the energy denominator $D_\circ$
in (17), where $D_\circ=p_\circ+k_\circ+k'_\circ-p'_\circ$. $q$ is
the momentum transferred by the gluon. We have followed our
arguments in section III to calculate the Dirac structure in (17).

When the diquark is vector, there are two configurations for
addition of spin angular momenta and moreover that for each case
there are contributions from chromoelectric and chromomagnetic
form factors. For production of the so called spin-1/2$'$ baryon
with chromoelectric form factor we have

\begin{widetext}
\begin{eqnarray}
T_{VE{1/2}'}&=&\frac{g_s^2C_{F}f_B }{2m_B^2\sqrt{6
p_{0}p'_{0}k_{0}k'_{0}}}\frac{F_E(Q^2)}{D_\circ q^2} \bigl\{
\overline{U}_B[\rlap /\chi' (\rlap /P +m_B)\gamma_5(\rlap /k'
+m_D)] u(p')\bigr\}.
\end{eqnarray}
Similar the amplitude with chromomagnetic form factor is put in
the following form
\begin{eqnarray}
T_{VM{1/2}'}&=&\frac{g_s^2C_{F}f_B }{2m_B^2\sqrt{6
p_{0}p'_{0}k_{0}k'_{0}}}\frac{F_M(Q^2)}{D_\circ q^2} \bigl\{
\overline{U}_B[ 2\beta(P.\chi')\gamma_5(\rlap /P
 -2m_B)+\rlap /k'\gamma_5(\rlap /P +m_B)\rlap /\chi']
 u(p')\bigr\}.
\end{eqnarray}
Likewise there are two contributions for spin $3/2$ baryon
formation. They are
\begin{eqnarray}
T_{VE{3/2}}&=&-\frac{g_s^2C_{F} f_B}{8m_B^2\sqrt{2
p_{0}p'_{0}k_{0}k'_{0}}}\frac{F_E(Q^2)}{D_\circ q^2} \bigl\{
\overline{\Psi}^\mu_{{3}/{2}}[\rlap/\chi' \gamma_\mu(\rlap /P
+m_B)(\rlap /k' +m_D)] u(p')\bigr\},
\end{eqnarray}
and
\begin{eqnarray}
T_{VM{3/2}}&=&\frac{g_s^2C_{F} f_B}{8m_B^2\sqrt{2
p_{0}p'_{0}k_{0}k'_{0}}}\frac{F_M(Q^2)}{D_\circ q^2} \bigl\{
\overline{\Psi}^\mu_{{3}/{2}}
[2\beta(P.\chi')(2P_\mu-m_B\gamma_\mu)+ \rlap /k'\gamma_\mu(\rlap
/P +m_B) \rlap/\chi']u(p')\bigr\}.
\end{eqnarray}
For a spin-3/2 baryon with four momentum $P$, summation over
helicity states is carried out by means of the following
projection operator [13]

\begin{eqnarray}
{\mathcal P}_{3/2}^{\mu\nu}(P)&=&
g^{\mu\nu}-\frac{1}{3}\gamma^\mu\gamma^\nu-\frac{1}{3P^2}
(\rlap/P\gamma^\mu P^\nu+P^\mu\gamma^\nu\rlap/P).
\end{eqnarray}

Next we calculate the fragmentation functions. For spin-1/2 baryon
including a scalar diquark we put (17) into (11). Performing
average/sum over intial/final spin states we find

\begin{eqnarray}
D_{1/2}(z,\mu_\circ)&=&\frac{1}{2}\biggl(\frac{\pi\alpha_sf_{B}C_{F}}{m_B}\biggr)^2
\int\frac{d^3{\bf p}d^3{\bf k}d^3{\bf k'}\delta^3({\bf p}+{\bf
k}+{\bf k'}- {\bf
p'})F_S^2(Q^2)}{D_\circ^2q^4p_{0}p'_{0}k_{0}k'_{0}}\nonumber\\
&&\times\bigl\{(p'.k')(P.k')+\beta(p'.k')+\alpha\beta(P.k')+\alpha\beta^2\bigr\}.
\end{eqnarray}
In the case of spin-1/2$'$ baryon, employing a vector diquark,  we
should add up the chromoelectric and chromomagnetic contributions
given by (18) and (19). In this way we conclude that
\begin{eqnarray}
D_{1/2'}(z,\mu_\circ)&=&\frac{2}{3}\biggl(\frac{2\pi\alpha_sf_{B}C_{F}}{m_Dm_B^2}\biggr)^2
\int\frac{d^3{\bf p}d^3{\bf k}d^3{\bf k'}\delta^3({\bf p}+{\bf
k}+{\bf k'}- {\bf
p'})F_E^2(Q^2)}{D_\circ^2q^4p_{0}p'_{0}k_{0}k'_{0}}\nonumber\\
&&\times\kappa(P.k')^2\bigl\{\kappa\bigl[(p'.k')(P.k')+\beta(p'.k')+\beta^2(P.p')
+\alpha\beta(P.k')\bigl]+\beta^2(P.p')-\alpha\beta^2\bigr\}.
\end{eqnarray}
Similarly for a spin-3/2 baryon we obtain
\begin{eqnarray}
D_{3/2}(z,\mu_\circ)&=&\frac{(\pi\alpha_sf_{B}C_{F})^2}{3m_Dm_B^5}
\int\frac{d^3{\bf p}d^3{\bf k}d^3{\bf k'}\delta^3({\bf p}+{\bf
k}+{\bf k'}- {\bf
p'})F_E^2(Q^2)}{D_\circ^2q^4p_{0}p'_{0}k_{0}k'_{0}}\nonumber\\
&&\times\bigl\{4\kappa
(1+\kappa)(p'.k')(P.k')^2+\beta(3\kappa^2+5\kappa-1)(p'.k')(P.k')
-\beta^2(2-\kappa)(p'.k')\nonumber\\
&&+2\kappa (1+\kappa)(P.p')(p.k')^3+\beta(3\kappa^2+\kappa-2)(P.p')(P.k')^2 -\beta^2(\kappa+4)(P.p')(p.k')
\nonumber\\
&&-3\beta^3(P.p')+6\alpha\kappa
(1+\kappa)(P.k')^3+3\alpha\beta(2\kappa^2+2\kappa-1)(P.k')^2
-6\alpha\beta^2(P.k')-3\alpha\beta^3\bigr\}.
\end{eqnarray}
We have set up our kinematics such that the dot products in (23),
(24) and (25), take the following form
\begin{eqnarray}
2P.k'/m_B^2&=&\frac{z}{1-z}(\beta^2+\gamma^2)+\frac{1-z}{z},\qquad
2P.p'/m_B^2=z(\alpha^2+\gamma^2)+{1}/{z},\nonumber\\
2p'.k'/m_B^2&=&\frac{1}{1-z}(\beta^2+\gamma^2)+(1-z)(\alpha^2+\gamma^2)-2\gamma^2,
\end{eqnarray}
where we have defined

\begin{eqnarray}
\alpha=\frac{m_Q}{m_B},\quad\beta=\frac{m_D}{m_B},\quad\gamma=\frac{k'_{T}}{m_B}.
\end{eqnarray}

Finally to obtain the explicit form of the fragmentation
functions, we perform the phase space integrations [14]. In this
way we find the fragmentation functions for spin-1/2, 1/2$'$ and
3/2 baryons as follows. For spin-1/2 case we have

\begin{eqnarray}
D_{1/2}(z,\mu_\circ)& = &\frac{\pi^3}{2}\biggl(\frac{\alpha_s
C_Ff_BQ_S^2}{ m_B^3}\biggr)^2F_1 (z),
\end{eqnarray}
where $F_1(z)$ is given by

\begin{eqnarray}
F_1(z)&=&\big\{z^4(1-z)^4[\beta^2-2\alpha\beta(-1+z)+\alpha^2(1-z)^2+\gamma^2z^2]\big\}\big /\nonumber\\
&&\bigl\{[1-(1-2\alpha\beta-2\beta^2)z-(2\alpha\beta+\beta^2-\gamma^2)z^2)]^2\nonumber\\
&&\times[1+2(-1+\beta)z+(1-2\beta+\beta^2+\gamma^2)z^2]^3\bigr\}.
\end{eqnarray}
Here we have redefined
$\gamma=\langle{k'_{T}}^2\rangle^{\frac{1}{2}}/m_B$. The
fragmentation function for spin-1/2$'$ case is obtained as

\begin{eqnarray}
D_{{1/2}'}(z,\mu_\circ)& = &\frac{\pi^3}{3}\biggl(\frac{4\alpha_s
C_Ff_BQ_V^4}{ \beta^2m_B^5}\biggr)^2\frac{F_2 (z)}{G(z)}.
\end{eqnarray}

Here the functions $F_2 (z)$ and $G(z)$ are given by

\begin{eqnarray}
F_2(z)=\kappa f_1(z)+\kappa^2f_2(z),
\end{eqnarray}
and
\begin{eqnarray}
G(z)&=&\big[1-(1-2\alpha\beta-2\beta^2)z-(2\alpha\beta+\beta^2-\gamma^2)z^2\big]^2\nonumber\\
&&\times\big[1-2(1-\beta)z+(1-2\beta+\beta^2+\gamma^2)z^2\big]^6,
\end{eqnarray}
where
\begin{eqnarray}
f_1(z)&=&2\beta^2z^4(1-z)^6\big(1-2\alpha
z+\alpha^2z^2+\gamma^2z^2\big)
\big[1-2z+(1+\beta^2+\gamma^2)z^2\big]^2,
\end{eqnarray}
\begin{eqnarray}
f_2(z)&=&z^4(1-z)^4\big[1-2 z+(1+\beta^2+\gamma^2)z^2\big]^2\nonumber\\
&&\times\big\{\beta^4 z^2+2\beta^3z(1-z)(1+\alpha z)
+\big[\alpha^2(1-z)^2+\gamma^2z^2\big]\big[1-2z+(1+\gamma^2)z^2\big]\nonumber\\
&&+\beta^2\big[3-6z+3(1+\alpha^2+\gamma^2)z^2-2(3\alpha^2+2\gamma^2)z^3+3(\alpha^2+\gamma^2)z^4\big]\nonumber\\
&&+2\beta(1-z)\big[\alpha^2z(1-z)^2+\gamma^2z^3+\alpha
(1-2z+(1+\gamma^2)z^2)\big]\big\}.
\end{eqnarray}
Finally for a spin-3/2 baryon we have

\begin{eqnarray}
D_{3/2}(z,\mu_\circ)& =
&\frac{\pi^3}{6\beta^3}\biggl(\frac{\alpha_s C_Ff_BQ_V^4}{
m_B^{5}}\biggr)^2\frac{F_3(z)}{G(z)}.
\end{eqnarray}
Here the function $F_3(z)$ reads as

\begin{eqnarray}
F_3(z)=f'_1(z)+\kappa f'_2(z)+\kappa^2f'_3(z),
\end{eqnarray}
where
\begin{eqnarray}
f'_1(z)&=&-2\beta
z^4(1-z)^6\big[1-2z+(1+\beta^2+\gamma^2)z^2\big]\nonumber\\
&&\times\big\{1-(2-3\alpha-4\beta)z
+\big[1+2\alpha^2-4\beta+2\beta^2-6\alpha(1-2\beta)+2\gamma^2\big]z^2\nonumber\\
&&-\big[4\alpha^2(1-\beta)+2(1-2\beta)\gamma^2-3\alpha(1-4\beta+\beta^2+\gamma^2)\big]z^3
+(\alpha^2+\gamma^2)(2-4\beta+\beta^2+\gamma^2)z^4\big\},
\end{eqnarray}
\begin{eqnarray}
f'_2(z)&=&z^4(1-z)^4\big[1-2z+(1+\beta^2+\gamma^2)z^2\big]\nonumber\\
&&\times\big\{\beta^3z^2(1-z)^2(11+\gamma^2z^2)+\beta^4z^2\big[2-2(1-\gamma^2)z^2+\gamma^2z^4\big]\nonumber\\
&&+\beta(1-z)^2\big[1-2z+(1+2\gamma^2)z^2-2\gamma^2z^3+\gamma^2(11+\gamma^2)z^4\big]\nonumber\\
&&+\gamma^2z^2\big[1-2(3-\gamma^2)z^2+8z^3-(3+2\gamma^2)z^4\big]
+6\alpha(1-z)\big[1-2(2-\beta)z\nonumber\\
&&+2(3-3\beta+\beta^2+\gamma^2)z^2-2(2\beta^2-\beta^3+2(1+\gamma^2)-\beta(3+\gamma^2))z^3
\nonumber\\
&&-(2\beta(1+\gamma^2)-2\beta^2(1+\gamma^2))z^4\big]+\alpha^2(1-z)^2\big[1-(6-11\beta-2\beta^2-2\gamma^2)z^2+2(4-11\beta-\beta^2)z^3\nonumber\\
&&-(3+2\gamma^2-\beta(11+\gamma^2))z^4\big]+\beta^2\big[1-2z+2\gamma^2z^2+2(1-\gamma^2)z^3-(1-6\gamma^2)z^4\big]\big\},
\end{eqnarray}
and
\begin{eqnarray}
f'_3(z)&=&z^4(1-z)^4\big[1-2z+(1+\beta^2+\gamma^2)z^2\big]\nonumber\\
&&\times\big\{3\beta^3z^2(1-z)^2(3+\gamma^2z^2)+\beta^4z^2\big[2-2(1-\gamma^2)z^2+\gamma^2z^4\big]\nonumber\\
&&+3\beta(1-z)^2\big[1-2z+(1+2\gamma^2)z^2-2\gamma^2z^3+\gamma^2(3+\gamma^2)z^4\big]\nonumber\\
&&+\gamma^2z^2\big[1-2(3-\gamma^2)z^2+8z^3-(3+2\gamma^2)z^4\big]
+6\alpha(1-z)\big[1-2(2-\beta)z+2(3-3\beta+\beta^2+\gamma^2)z^2\nonumber\\
&&-2(2\beta^2+2(1+\gamma^2)-\beta(3+\gamma^2))z^3
+(-2\beta(1+\gamma^2)+2\beta^2(1+\gamma^2))z^4\big]\nonumber\\
&&+\alpha^2(1-z)^2\big[1+(-6+9\beta+2\beta^2+2\gamma^2)z^2+(8-18\beta)z^3
-(3+2\gamma^2+2\beta^2(1-\gamma^2)-3\beta(3+\gamma^2))z^4\big]\nonumber\\
&&+\beta^2(1-2(3-\gamma^2)z^2+8z^3-3z^4-2\gamma^2(1-\gamma^2)z^6)\big\}.
\end{eqnarray}
\end{widetext}
It is important to note that in obtaining the above fragmentation
functions, we have used the scalar and vector form factors given
in (2). We have explained this in more detail in the following
section.

\section{Results and discussion}

The fragmentation functions given by (28), (30) and (35) are
obtained assuming three massive constituents for the final state
baryon in relativistic manner. They may be applied to different
baryonic states. Note that we have only considered the diquarks
and baryons in their ground states. While the fragmentation
function (28) provides the description of a spin-1/2 baryon with a
scalar diquark, (30) and (35) will demonstrate the production of
spin-1/2$'$ and spin-3/2 baryons with a vector diquark. For singly
heavy baryons one should consider a light diquark to attach to an
initial state heavy quark. For doubly heavy baryons the attachment
of a light-heavy diquark to a heavy quark should be considered. In
this case the diquark in its ground state is a scalar diquark.
Therefore only the fragmentation function (28) may by used to
describe spin-1/2 doubly heavy baryons.

\begin{table*}
\caption{The fragmentation probabilities (F.P.), the average
fragmentation parameter $\langle z\rangle$ and the cross sections
obtained from (28), (30) and (35) for production of the states
specified. The values of the parameters $Q_S$ and $Q_V$ are also
given. Total cross sections are calculated at the scale of
$\mu=2\mu_R$ [14].}\vskip .5cm
\begin{ruledtabular}
\begin{tabular}{c c c c c c c}
Process&$\langle z\rangle$&F.P.&$Q_S$($Q_V$)[GeV]&Cross section (LHC) [pb]\\
\hline
$b\rightarrow \Omega_{bbc}$ &0.603&$5.12\times 10^{-6}$&7.0&35\\
$c\rightarrow \Omega_{bcc}$ &0.535&$2.47\times 10^{-6}$&4.9&30\\
$c\rightarrow \Omega_{ccc}$ &0.526&$2.77\times 10^{-5}$&2.9&310\\
$c\rightarrow \Omega_{ccc}^*$ &0.549&$2.76\times 10^{-5}$&2.9&310\\
\end{tabular}
\end{ruledtabular}
\end{table*}

As a typical application of this model, we consider the situation
of triply heavy baryons and as a prototype example for application
of (28), we consider the fragmentation production of
$\Omega_{bbc}$ and $\Omega_{bcc}$ in a $b$ or a $c$ quark
fragmentation. Here the diquark is a colored $bc$ state. Using
this function we have obtained the behavior of the fragmentation
functions for these states. The behavior of the fragmentation
functions are shown in Fig. 2. The fragmentation probabilities as
well as the average fragmentation parameters are also calculated
using (28). These results are in fare agreement with similar
results ones in [16]. An interesting and useful example for
application of (30) and (35) is the situation of fragmentation
production of $\Omega_{ccc}$ baryon. According to our early
discussion, the vector $cc$ diquark will produce either a
spin-1/2$'$, $\Omega_{ccc}$, or a spin-3/2, $\Omega_{ccc}^*$,
state in a $c$ quark fragmentation. Figure (3) describes the
situation. Although the quantity $\langle z \rangle$ is a little
bit higher for the spin-3/2$'$ case, nevertheless the
fragmentation probabilities are almost the same. This would mean
that the cross sections for $\Omega_{ccc}$ and $\Omega_{ccc}^*$
would be the same at any colliding facility at least within the
framework of our study. This result is mainly because we have
ignored the Fermi motion inside the baryon and $m_B$ has been
taken the same for both  $\Omega_{ccc}$ and $\Omega_{ccc}^*$.
These results are comparable with those obtained in [14] and [15]
where triply heavy baryons are studied in detail using
perturbation theory and without the idea of diquarks. Table II
shows all physical observables which results from (30) and (35)
for these two states.

\begin{figure}
\centering\epsfig{file=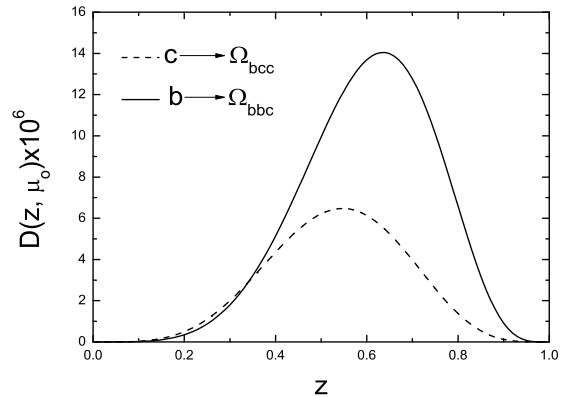,height=6.5cm,width=8.5cm,clip=} \hspace{1.5cm} %
\vspace{-.5cm} \caption[Submanagers]{ The fragmentation of $c$ and
$b$ quarks into $\Omega_{bcc}$ and $\Omega_{bbc}$ baryons
respectively. Here the diquark is a scalar $bc$. The curves are
plotted using (28).}
\end{figure}

\begin{figure}
\centering\epsfig{file=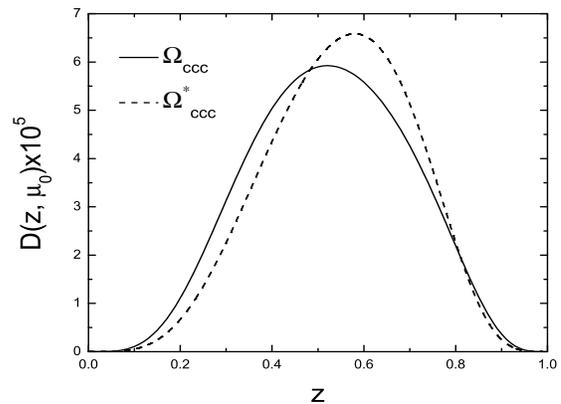,height=6.5cm,width=8.5cm,clip=} \hspace{1.5cm} %
\vspace{-.5cm} \caption[Submanagers]{ Fragmentation of a $c$ quark
into a spin-1/2, $\Omega_{ccc}$ or a spin-3/2, $\Omega_{ccc}^*$
baryon as obtained from (30) and (35) respectively. Here the
diquark is a vector $cc$.}
\end{figure}

For the fragmentation functions obtained in (28), (30) and (35) we
have used the input parameters as those employed in [14], whose
results are compared with those obtained here. The heavy quark
masses are $m_{b}=4.25$ GeV and $m_{c}=1.25 $ GeV. For all triply
heavy baryons we have chosen the decay constant to be $f_{B}=0.25$
GeV. The color factor for the color structure of the diagram in
Fig. 1 is $C_{F}={4}/({3\sqrt{3}})$. We have run the coupling
constant $\alpha_s$ according to appropriate momentum flow.

Application of the diquark form factors has introduced the
parameters $Q_S$ and $Q_V$ into our description for which there is
not much information at hand. Perturbative calculations of [14]
provide a clue. Although the fragmentation functions in the two
models are different, we assume that the fragmentation
probabilities provided by them for a given state are equal. This
procedure will fix the value of corresponding $Q_S$ ($Q_V$). The
fragmentation functions for production of $ \Omega_{ccc}$ and $
\Omega_{ccc}^*$ is shown in Fig. 4 from which we obtain the value
of $Q_V$ for $ \Omega_{ccc}$ and $ \Omega_{ccc}^*$ production.
Similarly the study of other processes will determine the
corresponding $Q_S$ ($Q_V$) for them. Some of the values for these
parameters are given in Table II. It is interesting to note that
the values of these parameters increase as the mass of the baryon
increases by inclusion of heavier and heavier flavors. The next
input parameter is the anomalous chromomagnetic dipole moment
$\kappa$. Different values have been attributed by different
authors for this quantity. Our study shows that among them the
value of $\kappa =1.39$ is more consistent with the behavior of
the fragmentation functions and physical quantities extracted from
them.

\begin{figure}
\centering\epsfig{file=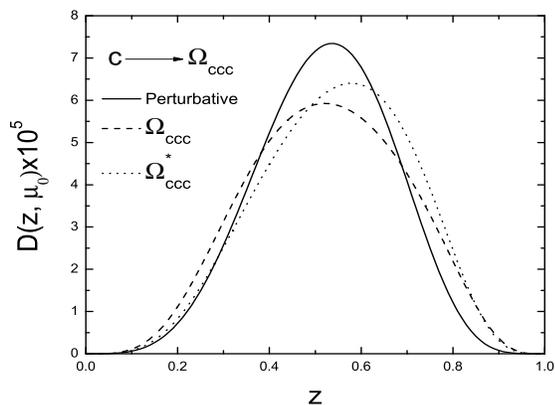,height=6.5cm,width=8.5cm,clip=} \hspace{1.5cm} %
\vspace{-.5cm} \caption[Submanagers]{ Obtaining $C_V$ for the
process of $c\rightarrow \Omega_{ccc}$ or $ \Omega_{ccc}^*$, from
comparison of the two models, the perturbative approach and
quark-diquark models. The fragmentation probabilities are taken to
be the equal.}
\end{figure}

At the end we would like to add that the ordinary parameterization
for the form factors describing the coupling of a gluon to heavy
diquark, i.e. the forms introduced in (1), just fail to provide
reasonable behavior of the fragmentation functions. Using them not
only leads to, unusual behavior of the fragmentation functions,
one needs abnormally large values of $Q_S$ and $Q_V$ to bring
about reasonable fragmentation probabilities. However the forms in
(2) look justified.

\end{document}